\documentclass{article}

\usepackage{arxiv}

\usepackage[utf8]{inputenc} 
\usepackage[T1]{fontenc}    
\usepackage{hyperref}       
\usepackage{url}            
\usepackage{booktabs}       
\usepackage{amsfonts}       
\usepackage{nicefrac}       
\usepackage{microtype}      
\usepackage{graphicx}
\usepackage{float}          

\graphicspath{ {./images/} }

\title{An IoT-Based Controlled Environment Storage for Prevention of Spoilage of Onion (Allium Cepa) During Post-Harvest with UV-C Disinfection}

\author{
  Shivam Kumar \\
  Department of Computer Science and Engineering (IoT \& IS)\\
  Manipal University Jaipur\\
  Rajasthan, India \\
  \texttt{shivam.23fe10cii00170@muj.manipal.edu} \\
  \And
  Himanshu Singh \\
  Department of Computer Science and Engineering\\
  Manipal University Jaipur\\
  Rajasthan, India \\
  \texttt{Himanshu.23fe10cse00124@muj.manipal.edu} \\
}

\begin{document}
\maketitle

\begin{abstract}
India is the second largest producer of onions in the world, contributing over 26 million tonnes annually. However, during storage, approximately 30--40\% of onions are lost due to rotting, sprouting, and weight loss. Despite being a major producer, conventional storage methods are either low-cost but ineffective (traditional storage with 40\% spoilage) or highly effective but prohibitively expensive for small farmers (cold storage). This paper presents a low-cost IoT-based smart onion storage system that monitors and automatically regulates environmental parameters including temperature, humidity, and spoilage gases using ESP32 microcontroller, DHT22 sensor, MQ-135 gas sensor, and UV-C disinfection technology. The proposed system aims to reduce onion spoilage to 15--20\% from the current 40--45\% wastage rate while remaining affordable for small and marginal farmers who constitute the majority in India. The system is designed to be cost-effective (estimated 60k--70k INR), energy-efficient, farmer-friendly, and solar-powered.
\end{abstract}


\section{Introduction}

Food security is a constant challenge globally, and post-harvest losses play a major role in substantial food waste. Onion is a staple crop of India, which directly affects prices and farmer livelihoods. The physiology of onions plays a major role in explaining why they are highly susceptible to spoilage during storage. The shelf life of onions primarily depends on two factors: Relative Humidity (RH) and Temperature. High humidity leads to fungal growth such as \textit{Aspergillus niger} (Black mold) and bacterial soft rot, while high temperatures lead to premature sprouting \cite{suravi2024}.

The major onion-producing states in India are Maharashtra, Madhya Pradesh, Karnataka, Gujarat, and Bihar. Maharashtra ranks first in onion production with a share of 35\%, followed by Madhya Pradesh with 17\% according to the 2023-24 report \cite{apeda2024}. Onion is a temperate crop but can be grown under a range of climatic conditions such as temperate, tropical, and humid environments. The best performance can be obtained in mild weather without extreme cold or heat \cite{onion_agriculture}.

Currently, there are two extreme solutions available. The first is traditional storage, which is low-cost but results in 30--40\% onion spoilage and rot. The second is cold storage, which extends the shelf life of onions but costs a large amount of money that is unaffordable for the majority of farmers \cite{grand_challenge}. This creates a significant gap between cost and efficacy.

This paper proposes the design and feasibility of a low-cost IoT-based model that uses different types of sensors and actuators to implement an automatically controlled storage system. We use DHT22 sensor for monitoring temperature and humidity, MQ-135 sensor for spoilage gas detection, and UV-C disinfection to limit chemical usage while preventing spoilage. The system is designed to be farmer-friendly, cost-effective, and solar-powered, extending the shelf life of onions without the large cost of cold storage.

The primary objectives of this paper are to: (1) present the complete architecture of the smart onion storage system, (2) detail the logic behind the technologies used, and (3) compare the economic impact of the prototype versus existing solutions.
\subsection{Qualitative Losses and Environmental Factors}
According to the ICAR Grand Challenge report, improper storage leads to significant quality degradation \cite{grand_challenge}.

\textbf{Qualitative losses:}
\begin{itemize}
    \item \textbf{Black mold:} 25--30\% reduction in market value
    \item \textbf{Outer skin removal:} 25--30\% reduction in market value
\end{itemize}

Various abiotic factors like temperature and relative humidity (RH) affect the health of onions; hence, their balance is needed to store the crop with minimum losses \cite{grand_challenge}:

\begin{itemize}
    \item High Temperature ($>32^{\circ}$C) + Low RH ($<60\%$) = \textbf{Weight loss}
    \item Low Temperature ($0-2^{\circ}$C) + High RH ($>70\%$) = \textbf{Sprouting}
    \item High Temperature ($>32^{\circ}$C) + High RH ($>70\%$) = \textbf{Rotting}
\end{itemize}
\section{Literature Review}

For the proposed solution, a detailed review of literature is essential to establish the scientific foundation. The idea revolves around the physiology of onions, environmental monitoring and control methods, and disinfection technologies.

\subsection{Physiology of Onion Spoilage (\textit{Allium cepa})}

According to research from \cite{suravi2024}, the two main causes of onion loss are sprouting and spoilage. High temperatures inside storage facilities cause early sprouting in onions, which reduces their market value. The second problem is spoilage due to fungal pathogens, especially \textit{Aspergillus niger} (black mold), which thrives in high humidity conditions generally above 75\% RH \cite{grand_challenge}. Among onion diseases, Fusarium alone causes 50\% of storage rot, which occurs mainly due to high temperatures and humidity \cite{suravi2024}. Therefore, storage requires both temperature and humidity management.

\subsection{Use of UV-C Disinfection}

Some fungal pathogens are resistant to antibiotics \cite{song2023}, leading to the use of harsh chemical products for prevention of molds during onion storage. A better alternative is to use UV-C light disinfectant for pathogens. Generally, UV-C light is harmful to skin, but in the range of 207--222 nm, it is safe for human exposure while being highly effective at killing harmful pathogens \cite{song2023}. UV-C light directly interferes with the DNA and RNA of microbes and fungi, breaking their genetic code and causing incorrect fusion, ultimately killing them and preventing further reproduction \cite{chen2020}. This approach can prevent the use of harsh chemicals while saving onions from fungal spoilage.

\subsection{Sensors in Agriculture}

Recently, there has been a drastic increase in the use of sensors in agriculture for monitoring humidity, temperature, soil moisture, and pest detection \cite{renke_sensors}. For our storage system, we use sensors for two different purposes: monitoring humidity and temperature, and detecting spoilage gases.

The DHT22 sensor monitors changes in temperature and humidity with a monitoring range of -40°C to +80°C and 0\% RH to 100\% RH \cite{renke_sensors}. The MQ-135 air quality sensor detects spoilage gases. When onion spoilage begins, onions release several volatile gases such as hydrogen sulfide, ammonia, and other sulfur compounds \cite{rutuja2025}. Real-time detection of these gases using the MQ-135 sensor allows us to observe sudden spikes in selected gases within the storage and take corrective actions by controlling humidity or activating UV-C disinfection.

\subsection{Combination of Two Studies}

While there is significant research on UV-C and IoT devices independently, no single study combines both approaches to provide a low-cost, efficient storage system for staples like onions.

\section{Methodology}

\subsection{System Architecture}

Our storage system architecture uses the following components:

\begin{itemize}
\item ESP32 (Wi-Fi + Bluetooth supported microcontroller)
\item DHT22 (Humidity and temperature sensor)
\item MQ-135 (Spoilage gas sensor)
\item UV-C lamp (Germicidal light 207--222 nm)
\item Dehumidifier (For reducing humidity inside storage)
\item Evaporative cooling pads
\item Industry-grade fans
\item LCD/OLED display (For real-time readings)
\item MQTT protocol (IoT connectivity)
\item Relay modules (For controlling high-voltage devices)
\end{itemize}

\begin{figure}[htbp]
  \centering
  \includegraphics[width=0.5\textwidth]{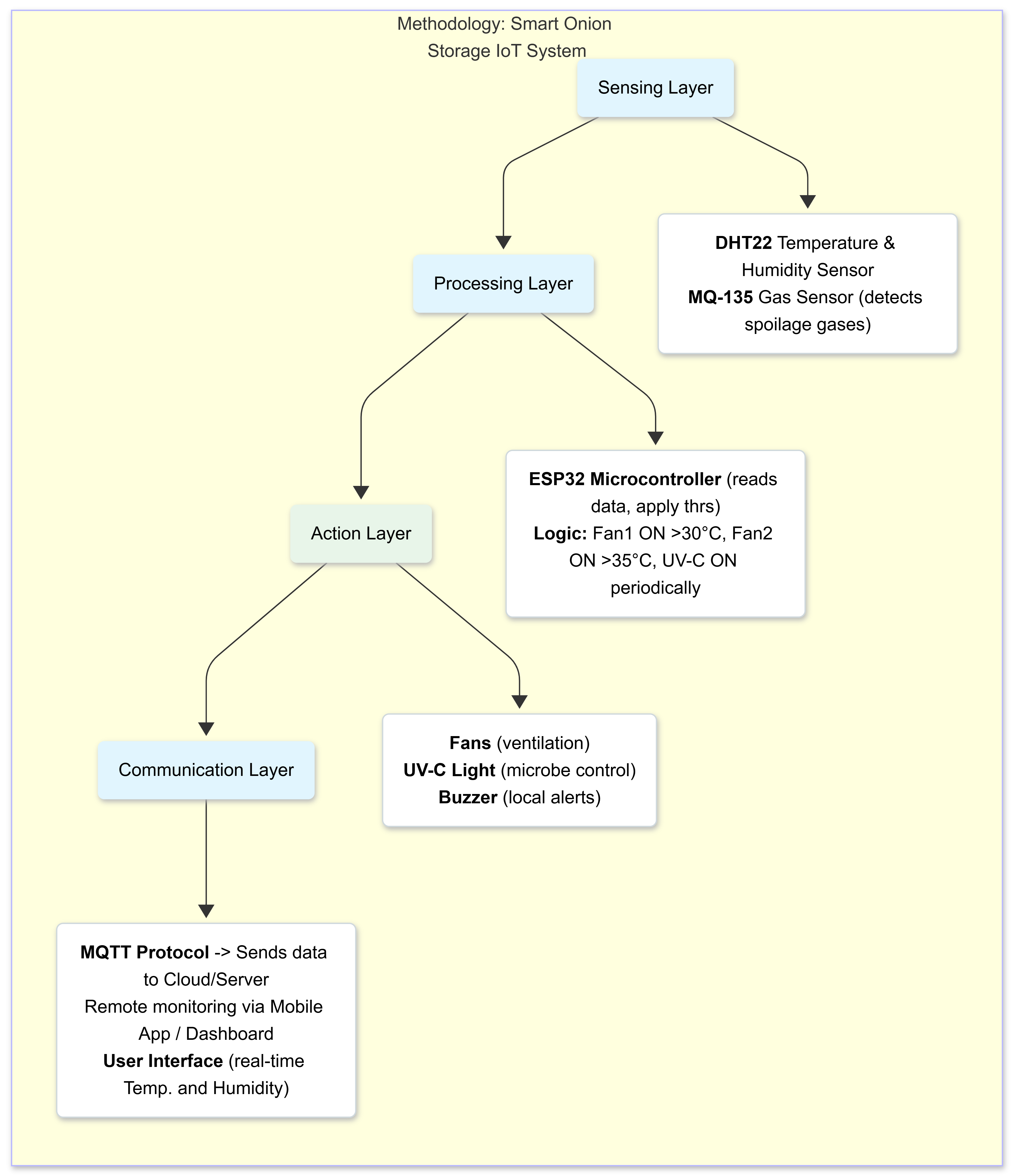}
  \caption{System Architecture Diagram}
  \label{fig:architecture}
\end{figure}

\subsection{Components Selection}

\paragraph{Microcontroller:}
We use ESP32 instead of Arduino Uno because of the integrated Wi-Fi and Bluetooth capabilities. This is essential for IoT connectivity with MQTT protocol. ESP32 also has a dual-core processor that enables simultaneous sensor reading and network communication \cite{esp32_arduino}. It supports multiple programming languages such as Python, C, and C++.

\paragraph{Sensors and Actuators:}
As discussed, the physiology of \textit{Allium cepa} (onion) makes it prone to spoilage due to temperature and humidity variations. We use the DHT22 sensor, which measures both parameters with high accuracy. We set sensor thresholds such that when temperature exceeds 30°C or humidity exceeds 75\% RH, the system triggers alarms or activates actuators. The MQ-135 air sensor detects spoilage gases, and when it reads a spike in ammonia and sulfide compounds, it signals the actuators.

In IoT systems, actuators are devices that convert electrical signals from sensors into physical actions such as motion or environmental changes. These are the "doers" while sensors are the "readers." Our actuators include the UV-C lamp (activated when sensors detect rot or high humidity), evaporative cooling pads, dehumidifiers, and industry-grade fans (triggered when humidity is high or temperature rises).

Normally, ESP32 only supports a maximum output of 5V, which is sufficient for small prototypes but inadequate for industry-grade fans, large dehumidifiers, and UV-C lamps in real warehouse-sized projects. Here, relay modules are employed. They work by taking an external power source as input along with the ESP32 output, using an electromagnet to open and close circuits to output devices \cite{relay_module}. This allows ESP32 to trigger high-voltage devices safely.

\subsection{Prototype of the Model}

Since we lack the required components for a real-life prototype, we created a simulation using Tinkercad (a free platform for students). Tinkercad does not support ESP32 and has limited components, so we made the following replacements:

\begin{itemize}
\item Arduino Uno instead of ESP32
\item Two DC motors instead of fans
\item Neon RGB lights instead of UV-C lamp
\item Potentiometer instead of DHT22
\item Buzzer instead of alarm
\item LCD display for real-time monitoring
\end{itemize}

\begin{figure}[htbp]
  \centering
  \includegraphics[width=0.95\textwidth]{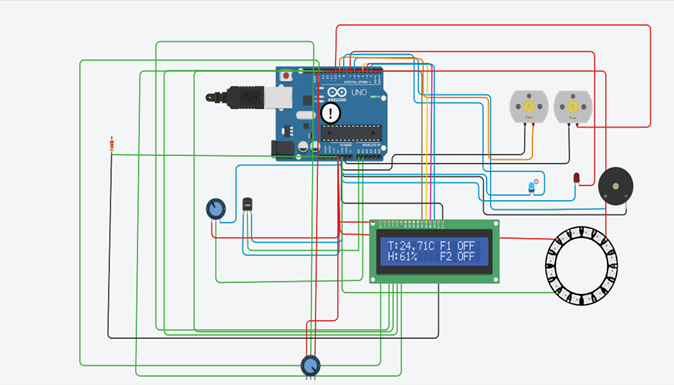}
  \caption{Circuit diagram of the simulated device}
  \label{fig:circuit}
\end{figure}

\begin{figure}[htbp]
  \centering
  \includegraphics[width=0.9\textwidth]{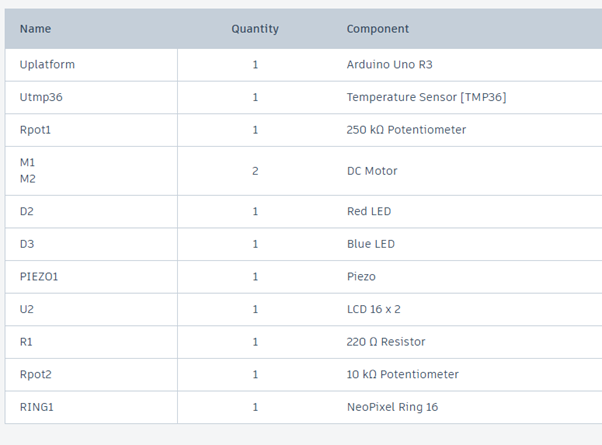}
  \caption{List of devices used in the Tinkercad simulation}
  \label{fig:devices}
\end{figure}

\paragraph{Functionality:}
Since we cannot have actual sensors inside Tinkercad to sense humidity, temperature, and gases, we used a potentiometer to adjust temperature and humidity levels. When we rotate the potentiometer clockwise, it increases the temperature, which triggers the fans to cool the atmosphere inside the storage.

\begin{figure}[htbp]
  \centering
  \includegraphics[width=0.9\textwidth]{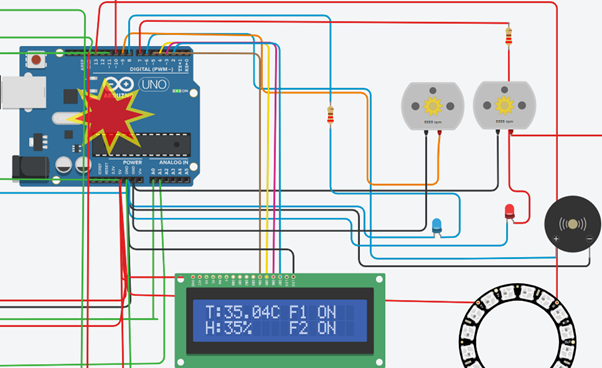}
  \caption{Fans triggered after temperature crosses the limit of 30°C}
  \label{fig:fans}
\end{figure}

Regarding fungal spread and black mold, these generally occur when humidity exceeds 70\% RH. To stop mold spread, we activate the UV-C light, which kills the fungal-causing pathogens.

\begin{figure}[htbp]
  \centering
  \includegraphics[width=0.7\textwidth]{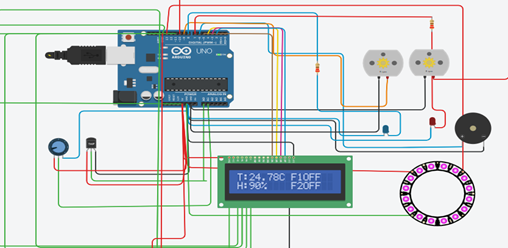}
  \caption{Activation of UV-C Lamp when humidity exceeds threshold}
  \label{fig:uvc}
\end{figure}

\paragraph{Core Logic Code Snippet:}
The Arduino code implements conditional logic to monitor sensor readings and trigger appropriate actuators based on threshold values.

\begin{figure}[htbp]
  \centering
  \includegraphics[width=0.9\textwidth]{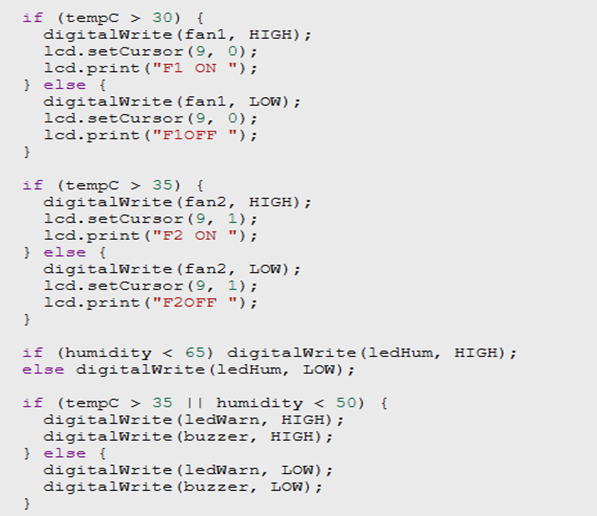}
  \caption{Arduino code snippet showing the control logic}
  \label{fig:code}
\end{figure}

\section{Analysis}

\subsection{Cost Effectiveness}

Compared to existing storage solutions that prevent onion spoilage, cold storage costs an astronomical amount for storing 10--50 metric tons of onions \cite{suravi2024}. India consists of 90\% small farmers who cannot afford such large sums. In contrast, our storage system has an estimated cost of 60,000--70,000 Indian rupees at large scale, which is approximately 10 times cheaper than cold storage while reducing onion spoilage to 15--20\% from the current 40--50\% rate.

\subsection{Health Benefits}

In other types of storage, harsh chemicals are used to prevent mold and fungus outbreaks, which affects consumer health. Our system uses a UV-C lamp in the safe range (207--222 nm), which directly kills fungi and stops mold spread without chemical residues.

\subsection{Limitations}

We are considering cost and prevention of spoilage based on existing research papers, as we have only simulated the technical components of our storage system and have not tested it in a real-life project. Field validation is required to confirm the effectiveness of the proposed system under actual storage conditions with varying onion varieties and environmental parameters.

\section{Conclusion}

In this paper, we have designed and implemented a smart storage system that is cost-effective and significantly better than existing cold storage and traditional storage solutions. This storage system is based on smart IoT devices such as ESP32, multiple sensors, and actuators including UV-C lamps and industry-grade fans. The system has the potential to reduce onion spoilage and sprouting during storage, decreasing the wastage rate by 20--25\% from the original wastage rate of 40--45\%.

The proposed system addresses the critical gap between expensive cold storage and ineffective traditional storage, making it accessible to small and marginal farmers who form the majority of India's agricultural workforce. Future work should include field trials with actual onion storage facilities, optimization of sensor thresholds for different onion varieties, and integration of machine learning algorithms for predictive maintenance and intelligent environmental control.

\bibliographystyle{unsrt}

\end{document}